\documentclass[12pt]{article}
\usepackage[margin=1in]{geometry}
\usepackage{amsmath,amssymb}
\usepackage{booktabs}
\usepackage{graphicx}
\usepackage{hyperref}
\usepackage{setspace}
\usepackage{enumitem}
\usepackage{xcolor}
\usepackage[normalem]{ulem}
\title{Verify and Protect Before You Trust: A Practical Protocol for  AI\\
and Algorithmic Adoption in Methodology Development\\
for National Statistical Offices}
\author{Siu-Ming Tam \\ Tam Data Advisory Pty Ltd}
\date{}
\begin{document}
\maketitle

\begin{abstract}
\noindent To meet increasing user demands for granular demographic and socio-economic indicators under tightening budgets, national statistical offices have to continuously engage in methodology research and development, including harnessing big data, satellite imagery, and transactional sources to improve or redesign data collection instruments. AI can help with that ongoing effort. This paper focuses
on using AI-generated algorithms to test, verify, and illustrate a new statistical methodology before it is trusted for production. Trust in the use of AI in official statistics has many dimensions;
we deliberately restrict this paper to two pillars: an independent statistical verification step before the algorithm is trusted for production; and disciplined protection of respondent confidentiality throughout development and testing. We illustrate this process of building trust, drawing on the author's own experience directing AI to build and run the algorithm for a Mini-Max Hierarchical Bayes (HB) sampling method. On a synthetic labour-force population the method met all stated precision targets with an 80 per cent reduction in required sample size, confirmed by a 1,000-replication Monte Carlo study; applied to real 2021 Australian Census microdata, it achieved a 90 per cent reduction with national point estimates accurate to well under 1 per cent. These figures are the outcome of one case study of one sampling method on one population, and are reported to illustrate the protocol at work, not as a general claim about the savings any AI-assisted method will deliver; transferring them to another method or setting requires repeating the same independent verification, not assuming the result. This paper also gives a checklist that any national statistical office can apply when evaluating an AI-assisted or algorithmic method for production use. The protocol and checklist are cross-checked against two established reference points for methodological practice in official statistics: the UN Fundamental Principles of Official Statistics and the HLG-MOS Quality Framework for Statistical Algorithms.
\end{abstract}

\noindent\textbf{Keywords:} AI Assistant; AI Protocol; Audit; Protect; Verify. 

\section{Introduction}
National statistical offices increasingly need smaller, cheaper, and more
granular estimates on demographic and socio-economic indicators to meet user demands under tightening
budgets. Meeting that need is not a one-off exercise but a continuous methodology
research and development effort, because the opportunities and pressures
driving it keep changing: harnessing big data from administrative sources,
satellite imagery, and transactional data as complements to survey
collection; a steadily increasing need for small area estimates as users
demand finer disaggregation; and survey operating budgets that are being
reduced rather than merely held flat.

National statistical offices have been living with this pressure for over a
decade. In 2014, the author set out a Methodology Architecture roadmap for
the Australian Bureau of Statistics diagnosing the same pressures under
different names: a ``products vision'' requiring the ability to combine
survey data with ``administrative data, transactional data, Big Data, and
`organic' data'', and a ``process vision'' requiring statistical methods and
tools to be innovated, industrialised, contemporised, and supported by
building capability and building support across the organisation
\cite{tam2014ma}. More than a decade on, the pressures are recognisably the
same ones: new data sources, tighter budgets, and a widening gap between what
is asked of national statistical offices and what classical survey methods
alone can deliver, arriving under a different name (``responsible AI''
rather than ``methodology architecture'').

This paper is about one specific way AI can help with that ongoing
methodology development: using AI-generated algorithms to test, verify, and
illustrate a new statistical methodology before it is trusted for
production. This is deliberately a narrower claim than ``responsible AI in
official statistics'' as a whole. Responsible AI use has many dimensions, including fairness across population groups, transparency of automated decisions, and ongoing monitoring after deployment, which are outside the scope of this paper. Here we restrict attention to a question that sits upstream of all of them: how AI's role in developing and testing a statistical method is kept
honest before that method is trusted and used in production at all.

AI is being taken up at other stages of the survey life cycle too. Kreuter \cite{kreuter2025}
surveys how large language models are already assisting with questionnaire
design, including drafting candidate items, translating instruments, and
flagging inconsistent wording, alongside their emerging role in interviewing,
data processing, and analysis. \emph{Journal of Official Statistics} devoted
its 40th-anniversary special issue (volume 41, issue 3, 2025) to exactly this,
covering machine learning for estimation and small area estimation \cite{tzavidis2025,ranalli2025}, statistical disclosure control and privacy-enhancing technologies \cite{domingoferrer2025}, and the
broader ethical and data-collection implications of these changes. Most of
that issue, agenda-setting by design, asks what the field should work on
next. This paper's contribution is narrower and more operational: a worked
protocol and checklist a methodologist can apply to a specific AI-assisted
methodology development. It also provides a real-life illustration of how the protocol and checklist were applied to develop the Mini-Max HB sampling methodology to reduce survey operating costs.

This paper's two pillars should be read alongside, not in place of, the broader governance frameworks already emerging for AI in official statistics. UNECE's Framework on Responsible AI for Official Statistics \cite{unece2025}, developed by the Applying Data Science and Modern Methods Group under the High-Level Group for the Modernisation of Official Statistics, sets out six principles for AI use across the whole statistical production cycle: ethical purpose and public good, accountability, transparency and explainability, fairness and non-discrimination, privacy and security, and validity and robustness. That framework operates at the level of principles, deliberately spanning every use of AI in an NSO, from public-facing chatbots to automated coding to methodology development. It does not, and is not designed to, tell a methodologist what ``validity and robustness'' looks like in the specific moment an AI assistant has just returned code implementing a statistical method and the question is whether to trust it. This paper's contribution sits at exactly that narrower point: a worked, reusable operational procedure for the validity-and-robustness and privacy-and-security principles specifically in AI-assisted statistical methodology development, illustrated end to end, including where it did not go as planned, on a real design problem. The checklist in Section~6 is offered as one operational instance of those two principles, not as a substitute for the fuller framework.

The Sixth UN World Data Forum, to be held in Riyadh in November 2026, includes harnessing AI responsibly as one of its programme tracks; that track is the occasion for this
paper, not its subject. A common but, we argue, incomplete answer to the
responsibility question focuses on provenance: who wrote the code, and how
much human oversight was applied at each step. Provenance matters, but on its own it is not what makes a statistical method safe to deploy or safe for the
people whose data it uses. Two further things do, and they form this paper's
two pillars. The first is evidence that the method does what it claims, that
its credible intervals cover the true value at the stated rate and its point
estimates are not systematically biased, established precisely by using
AI-generated algorithms to simulate data and test the method against a known
truth. The second is evidence that developing and testing the
method never required exposing confidential respondent data to unnecessary
risk.

The layout of the paper is as follows: Section 2 outlines a verification protocol for AI-generated algorithms. Section 3 describes the protocol to protect individuals' data. Section 4 cross-checks the protocol against the Fundamental Principles of Official Statistics, and Section 5 cross-checks it against the HLG-MOS Quality Framework for Statistical Algorithms. Section 6 outlines a checklist to be used to ensure AI-generated algorithms can be trusted. Section 7 describes the case study used to develop the Mini-Max HB sampling methodology, and Section 8 concludes.

\section{A Verification Protocol: The First Pillar}
The 2014 methodology architecture road map diagnosed, specifically, why model-based
methods were not moving into production alongside design-based ones at the
Australian Bureau of Statistics: they were, and largely still are, ``seldom
used in the collection and processing phases of the [statistical production
cycle], primarily because of concerns about quality of the estimates with
model mis-specification, and the effort required to develop, test and
evaluate models'' \cite{tam2014ma}. That sentence names a real, specific
reason: not that testing is impossible, but that building and running enough
of it to earn trust in a model-based method takes labour that a small
methodology team must find room for alongside everything else it is asked to
do.

Simulating a method against known truth, checking it against independent
benchmarks, and iterating until it holds up has been the gold standard for developing new
statistical methods. The cost of performing the iterative steps in methodology development in an NSO is high. However, an AI assistant changes the cost of meeting that gold
standard because work that used to require substantial
hand-written programming can now be done, provided it is under the direction of a methodologist, faster and more
thoroughly than before.

``Verification'' in this paper is used deliberately to cover three related but distinct questions, and the checklist in Section~6 is aimed squarely at the first two: (a) verification of the statistical method itself, that its estimator and its stated precision or coverage properties hold under the data-generating process it was designed for; (b) verification of the software implementation, that the code an AI assistant produced actually computes what the method specifies, rather than something that merely looks plausible; and (c) validation of fitness for operational production, that the method and its implementation continue to perform once embedded in a live collection with its own data flows, timelines, and edge cases. This paper's protocol and case study concentrate on (a) and (b), which is where an AI assistant does most of the work and where the correlated-error risk of Section~2.2 is sharpest. Validation in sense (c) is a further, largely institutional step, informed by but not reducible to (a) and (b), and is discussed only briefly, in the Conclusion, as a necessary sequel rather than as something this paper's protocol itself delivers.

\subsection{Three Roles in AI-Assisted Methodology Development}
The protocol rests on a division of labour no differently from
how a research director works with a capable research assistant in any
university department. Three roles need to be kept distinct:
\begin{itemize}
\item \textbf{The methodologist} develops the method and specifies the statistical targets: the
  estimator, the design objective, the precision or coverage criteria a
  method must meet, and the scenarios under which it will be tested. This is
  an intellectual and professional judgement that cannot be delegated.
\item \textbf{The AI assistant} may help with implementation, including
  writing code, generating simulated populations with a known ground truth,
  and running large numbers of Monte Carlo replications, under the
  methodologist's direction.
\item \textbf{The auditor} is the arbiter. Nothing produced by
  the second role is adopted for production use until it has been checked
  against something the assistant did not itself generate: a published
  benchmark, real historical survey data, or a hand-verified calculation.
\end{itemize}

Table~\ref{tab:roles} summarises the three roles alongside their primary responsibility and the minimum degree of independence each requires from the others.

\begin{table}[h]
\centering
\small
\begin{tabular}{p{0.18\textwidth}p{0.38\textwidth}p{0.34\textwidth}}
\toprule
\textbf{Role} & \textbf{Primary responsibility} & \textbf{Required independence} \\
\midrule
Methodologist & Sets the estimator, design objective, and acceptance thresholds before testing begins. & Independent judgement; not delegated to the AI assistant. \\
AI assistant & Implements the method: writes code, generates simulated populations, runs replications, under the methodologist's direction. & None required of the assistant itself; it works under direction and is not expected to be independent of the methodologist. \\
Auditor & Checks the assistant's output against something the assistant did not itself generate, before the method is adopted for production. & Must not share the assistant's own output as its reference point (see below for what counts as sufficient independence). \\
\bottomrule
\end{tabular}
\caption{Roles, responsibilities, and required independence in AI-assisted methodology development.}
\label{tab:roles}
\end{table}

The auditor role is the one most often skipped, and skipping it is what
turns faster AI-assisted development into greater risk rather than greater
trust.

Note that the same efficiency gain extends to the
auditor role as well: a second, independently configured AI system auditing
or cross-checking the first assistant's implementation before it is relied
upon, an automated analogue of independent code review, alongside rather
than instead of the checks described in the next subsections. Indeed, Zeng et al. \cite{zeng2025airepr} describe data-analysis workflows that use a second AI system to review code written by the first.

What counts as ``independent'' for the auditor role is not all-or-nothing; it is closer to a hierarchy of increasing strength. The weakest form, a fresh session of the same AI system, prompted separately and without sight of the first session's reasoning, already catches the class of errors that arise from one session simply not re-checking its own output, and is better than no audit at all, but it shares the same model, training data, and systematic blind spots as the assistant it is checking, so it cannot catch an error the model is disposed to make consistently. A stronger form uses a different AI provider or model family as the auditor, which removes shared model-level blind spots but still leaves a shared reliance on AI-generated reasoning as such. The strongest and most reliable form anchors the check outside AI entirely: a differently implemented benchmark (a different language, library, or algorithm reaching the same target, as in the fresh-draw validation reported under Item~3 of Section~7), a human reviewer with no role in development, or real data with an independently known answer. The auditor role in Section~2.1 requires, at minimum, the second of these three levels; the case study in Section~7 reports which level was actually achieved at each step, rather than presenting all checks as equally independent.

\subsection{Why independence matters: the correlated-error risk}
Why does the algorithm verification and building process have to be independent? This is because if the same tool is used to
build both the method and its verification, a shared error can appear in both processes and silently cancel out,
producing a result that appears to pass when the underlying method is in
fact wrong. The mitigation is to
anchor the verification against something built
independently of the method under test, such as a published result, a real
historical dataset with known properties, or an arithmetic check performed by
hand. A verification study that only checks itself against itself is not
verification, but one that is checked against an external, independently derived process or
benchmark is.

\section{A Second Pillar: Protecting Respondent Data}
Statistical verification answers whether a method and its associated algorithm work. It does not by
itself answer whether the method was developed and tested in a way that
protects the people whose data it draws on, a distinct and equally
necessary dimension of responsible AI use in official statistics,
particularly once an AI assistant is doing part of the underlying
computation. This is particularly important when the methodology development is carried out outside the NSO.

Test populations for verification are, wherever possible, synthetic or built
from published aggregate statistics rather than real individual-level
microdata, precisely so that the verification testing itself never requires
exposing confidential respondent records to any tool, AI-assisted or
otherwise. Synthetic data is a safeguard, not a free substitute for the real thing: a synthetic population is only as good as the assumptions used to build it, and it can systematically fail to reproduce rare subpopulations, complex dependencies between variables, and the operational anomalies (partial responses, coding errors, outliers) that real collections actually contain. A method that clears every target on synthetic data has been protected from confidentiality risk, not from this gap; closing it, where it matters, is precisely why Section~7's case study also checks the method against real microdata under governed access, rather than relying on synthetic testing alone. Where an agency holds real historical microdata and wishes to
supply it to an external methodologist as a gold-standard input, because no simulation assumption is then
needed, that remains a case-by-case exception handled under its own
data-handling agreement, not a default, and the computation involved is
carried out under the same controls that would apply to any other
confidential statistical processing, regardless of which tool performs it.

In practice, these controls map onto the 5 Safes framework increasingly used by NSOs to manage disclosure risk in data access: safe projects, safe people, safe settings, safe data, and safe outputs, under which risk is managed jointly and severally across all five dimensions rather than by relying on any one of them \cite{tam2021fivesafes}. As a researcher operating under this protocol, we owe the same respect to all five: a safe project (the verification task itself, scoped and agreed in advance, rather than an open-ended exploration of the data); safe people (only those named in the data-handling agreement); safe data (the extract minimised to the variables actually needed); and safe outputs (results cleared for disclosure risk before they leave the environment).

\cite{tam2021fivesafes} shows that, provided two independence assumptions hold across the five controls --- that satisfying each safe is statistically independent of satisfying the other four (unconditional independence), and that, given a disclosure has in fact occurred, which control is responsible is likewise independent of the others (conditional independence) --- the probability of disclosure given all five safes is the product of the five individual conditional disclosure probabilities, divided by the unconditional disclosure probability (with no controls at all) raised to the fourth power: $\Pr(D\mid S_1\cap S_2\cap S_3\cap S_4\cap S_5) = \Pr(D\mid S_1)\,\Pr(D\mid S_2)\,\Pr(D\mid S_3)\,\Pr(D\mid S_4)\,\Pr(D\mid S_5)\,/\,\Pr(D)^4$, where $S_1,\dots,S_5$ denote the project, people, settings, data, and output controls respectively.

Two practical implications follow directly from this product structure. First, because the overall probability is a product of five conditional probabilities, a very small conditional probability on any single dimension pulls the overall probability down further still, whatever the other four look like: this is why releasing aggregate rather than unit-record data carries a small overall disclosure risk even when the project, people, settings, and output controls are only ordinary, since aggregation itself typically gives the safe-data dimension a small conditional probability by construction (individual respondents are generally not identifiable from an aggregate). Second, the same route is available for individual-level microdata, which lacks aggregation's built-in advantage: tightly controlling the environment (safe settings) or the extract itself (safe data), for instance through an accredited secure research environment or heavy confidentialisation, can drive that dimension's own conditional probability down to a comparably small level and thereby achieve similarly strong overall protection by a different route.

For an AI-assisted workflow specifically, safe settings deserves particular emphasis: it is the one control a methodologist working outside the NSO, as this protocol assumes, is directly responsible for, rather than merely a beneficiary of controls the NSO has already put in place elsewhere. Before releasing any data, an NSO assesses the proposed use against all five safes and decides accordingly what form the release takes. Some releases are aggregate statistics or highly collapsed classificatory tables; others are public-use microdata files, typically a small random sample of the master file with classificatory detail heavily collapsed to reduce identifiability; and others are unit-record extracts released for a named project under a governed access arrangement, which some NSOs condition on the researcher completing confidentiality training and signing a data-handling agreement. Whatever form the release takes, the methodologist's role is to adhere to the conditions attached to it, not to independently reconstruct equivalent protections. For an AI-assisted workflow, that includes keeping the AI assistant within whatever environment those conditions specify: an assistant that is allowed to operate outside it, for instance by sending microdata to an externally hosted service, breaches the conditions of release regardless of how strong the other four controls are. This is not a novel requirement; it is the standard condition of any governed data release, restated here because an AI-assisted workflow is exactly the setting in which it is easiest to reach for a convenient external tool without noticing that a condition of release has been breached.

\section{Checking the Protocol Against the Fundamental Principles of Official Statistics}
A natural question is how the two-pillar protocol set out above complies with the United Nations Fundamental
Principles of Official Statistics (FPOS). FPOS predates AI entirely, which is
itself informative: if a verification-and-confidentiality discipline
built for an AI-assisted workflow turns out to line up with a
30-year-old, technology-neutral statement of professional practice, that
gives assurance that the protocol aligns with best practice for official statistics.
Table~\ref{tab:fpos} checks the protocol against all ten principles, presented as an interpretive crosswalk rather than a claim of formal compliance. It can be seen that three principles are directly and substantively engaged by
the two pillars; a further four are adjacent, in the sense that the
protocol's practice serves the same underlying concern without being a
complete implementation of the principle; three principles sit outside
the protocol's scope entirely: they concern legal frameworks, national
coordination, and dissemination policy.

\begin{table}[h]
\centering
\small
\begin{tabular}{p{0.24\textwidth}p{0.1\textwidth}p{0.58\textwidth}}
\toprule
\textbf{FPOS Principle} & \textbf{Fit} & \textbf{Note} \\
\midrule
1. Relevance, impartiality, equal access & \textbf{Outside scope} & Concerns dissemination policy and user access, not method verification. \\
2. Professional standards, scientific principles, professional ethics & \textbf{Direct} & The checklist \emph{is} an operationalisation of this principle: methods and procedures decided on stated, professional/scientific grounds rather than after the fact. \\
3. Accountability and transparency & \textbf{Direct} & Checklist item 3 (disclose shortfalls rather than remove them from the test suite) and the practice of retaining dated test logs and fixtures directly serve this principle. \\
4. Prevention of misuse & \textbf{Adjacent} & Disclosing a known limitation in advance (e.g.\ the domain-coverage shortfall in Section~7) helps forestall misinterpretation, but the principle is mainly about post-publication correction of others' misuse, which this protocol does not itself cover. \\
5. Sources of official statistics & \textbf{Adjacent} & The confidentiality pillar's preference for synthetic/aggregate test populations parallels this principle's cost/quality/burden reasoning, but Principle 5 concerns the choice of \emph{production} data sources, not test data used in method development. \\
6. Confidentiality & \textbf{Direct} & Section~3 is a direct application: real microdata treated as a governed exception, not a default, under the same controls regardless of which tool performs the computation. \\
7. Legislation & \textbf{Outside scope} & A legal/institutional matter for the agency, not the method-development protocol. \\
8. National coordination & \textbf{Outside scope} & Concerns coordination across agencies within a country; not addressed here. \\
9. Use of international standards & \textbf{Adjacent} & Checklist item 2's requirement to check against an independent benchmark is easiest to satisfy when that benchmark is a recognised method (Bethel allocation, Fay--Herriot, Prasad--Rao), which is a narrow echo of this principle's concern with common standards, not a full implementation of it. \\
10. International cooperation & \textbf{Adjacent} & This paper's own purpose, offering a transferable checklist to other national statistical offices, is a small instance of the cooperation this principle describes. \\
\bottomrule
\end{tabular}
\caption{The two-pillar protocol checked against the ten UN Fundamental Principles of Official Statistics.}
\label{tab:fpos}
\end{table}

\section{Checking the Protocol Against the Quality Framework for Statistical Algorithms}
FPOS is a technology-neutral statement of professional practice; a second, more directly relevant reference point is the Quality Framework for Statistical Algorithms (QF4SA) \cite{yung2022}, developed under the same HLG-MOS umbrella referenced in Section~1 specifically to address the quality implications of machine learning and other modern algorithms in official statistics. QF4SA proposes five quality dimensions for a statistical algorithm: accuracy, explainability, reproducibility, timeliness, and cost effectiveness. QF4SA was conceived with algorithms that produce \emph{intermediate} statistical outputs in mind, such as classification or imputation, rather than with AI-assisted development of the method itself; even so, its five dimensions travel well onto the protocol set out here, and checking against them is a useful complement to Table~\ref{tab:fpos}, because QF4SA is aimed squarely at the choice and evaluation of statistical algorithms, which is this paper's subject, in a way FPOS's ten principles are not.

Table~\ref{tab:qf4sa} sets out the correspondence. Two of QF4SA's five dimensions, accuracy and reproducibility, are directly and substantively engaged by the protocol's checklist; timeliness is directly engaged by the case study's empirical observations about AI-assisted development speed; explainability and cost effectiveness are adjacent, in the sense that the protocol touches the same underlying concern without QF4SA's own machinery (feature-importance methods, an accuracy-per-unit-cost metric) being adopted here.

\begin{table}[p]
\centering
\footnotesize
\begin{tabular}{p{0.2\textwidth}p{0.12\textwidth}p{0.6\textwidth}}
\toprule
\textbf{QF4SA Dimension} & \textbf{Fit} & \textbf{Note} \\
\midrule
Accuracy & \textbf{Direct} & QF4SA defines accuracy as the closeness of computations or estimates to the true value, distinguishing unit-level (G1), distributional (G2), and target-parameter (G3) accuracy. Checklist items 1, 2, and 5 (Section~6) are a direct operationalisation of this dimension for the Mini-Max HB method: the case study's CV targets, ARE and MARE thresholds, and repeated-sampling credible-interval coverage checks (Section~7) are exactly QF4SA's accuracy indicators applied end to end to a specific statistical algorithm. \\
Explainability & \textbf{Adjacent} & QF4SA's explainability concerns the relationship between an algorithm's inputs and outputs, a concern sharpest for opaque ML methods. The Mini-Max HB method is a parametric hierarchical Bayes model with an explicit functional form, so explainability in QF4SA's sense is largely satisfied by the method's own transparency rather than by anything this protocol adds. This paper's own explainability concern is different in kind: understanding what an AI assistant did to the code, which checklist item 4 and the auditor role (Section~2.1) address, but not in QF4SA's input-output sense. \\
Reproducibility & \textbf{Direct} & QF4SA recommends NSOs adopt methods reproducibility (replicating results with the same data, tools, and an arm's-length second analyst) and inferential reproducibility (corroborating results with different but applicable algorithms or assumptions), and explicitly sets aside results reproducibility as infeasible for NSOs. Checklist item 4's independent re-running of the verification script, and item 7's dated record, are direct instances of methods reproducibility. The fresh, independently implemented validation reported in Section~7 (a different language and library, a new random seed, corroborating rather than duplicating the original result) is a worked example of inferential reproducibility in exactly the sense QF4SA recommends. \\
Timeliness & \textbf{Direct} & QF4SA recommends adding development and processing time to the conventional timeliness concept, on the grounds that modern algorithms can shorten both. The case study's observation that AI-assisted diagnostic scripting ``took minutes rather than the days the same sequence of checks would otherwise have needed'' (Section~7, item 3) is a direct empirical instance of this dimension, with the verification standard itself held constant rather than relaxed to achieve the speed-up. \\
Cost effectiveness & \textbf{Adjacent} & QF4SA defines cost effectiveness as accuracy per unit cost, weighing fixed costs (infrastructure, training) against ongoing costs. This paper's own motivation, a smaller sample at equal precision, is a downstream cost saving of the kind QF4SA is concerned with, and the assemble-to-order discussion (Section~7, item 4) touches the same fixed-cost logic QF4SA sets out for adopting new tooling or components, but this paper does not compute an accuracy-per-unit-cost metric in QF4SA's own terms. \\
\bottomrule
\end{tabular}
\caption{The protocol and checklist checked against the five QF4SA quality dimensions.}
\label{tab:qf4sa}
\end{table}

\section{A Protocol Checklist}
The following checklist is proposed for use by NSOs when evaluating an AI-assisted or algorithmic method for official statistics production. Items 1--5 operationalise the verification pillar of Section~2; item 6 operationalises the confidentiality-protection pillar of Section~3; item 7 documents both.

To be explicit about what kind of instrument this is: the checklist is a minimum baseline, not a formal assessment or certification tool. It states the smallest set of things that should be true before an AI-assisted method is trusted for production; meeting all seven items is necessary but is not, on its own, a guarantee of fitness for a specific operational use, which will often call for additional agency-specific requirements (governance sign-off, legal review, production-system testing) on top of it. It is offered as a starting point a methodology team can adopt directly or adapt, not as a rigid audit standard to be scored against.

\subsection{Verification (Pillar 1)}
\begin{enumerate}
\item State the target statistical property and the acceptance threshold
  \emph{before} testing, not after.
\item Test by simulation against a data-generating process with known
  ground truth, whether synthetic, real historical data, or both, using
  enough replications that a single noisy draw cannot be mistaken for a
  genuine result.
\item Where a test falls short, iterate: refine the method and retest
  before concluding anything; where the gap still does not close, disclose
  the limitation rather than narrowing the test suite until it disappears.
\item Re-apply the same standard to the verification tooling itself, not
  only to the method it is testing.
\item Test at the operating scale the method will actually see in
  production, not only at the scale most convenient for a first check.
\end{enumerate}

\subsection{Confidentiality Protection (Pillar 2)}
\begin{enumerate}[resume]
\item Prefer synthetic or aggregate-derived data for testing over real
  individual-level microdata; treat a request for real microdata as a
  governed, case-by-case exception rather than a default. Accept whatever the
  NSO's own 5 Safes assessment (Section~3) determines is the appropriate
  response to that request, which may still be a public-use file with
  classificatory detail heavily collapsed rather than the underlying
  unit-record extract; where a unit-record extract is supplied, adhere strictly
  to the conditions attached to it, which may include confidentiality training
  and a signed data-handling agreement specific to that use. For an
  AI-assisted workflow, this means keeping the AI assistant within whatever
  environment those conditions specify, and never extending it access the
  release conditions do not cover.
\end{enumerate}

\subsection{Documentation (Supporting Both Pillars)}
\begin{enumerate}[resume]
\item Keep a dated record of what was tested, against what, and with what
  result; the record is as much a safeguard for institutional continuity
  as it is evidence of correctness. This record should include version control of the code, the fitted model objects, the AI prompts used to generate or revise them, the package and library versions the computation depended on, and the data inputs (or, for confidential data, a description sufficient to identify the exact extract used); an undated script or an unpinned package version is not a reproducible record even when the result it once produced was correct, because a later re-run under a different package version or a since-edited prompt is not the same test.
\end{enumerate}

\section{An example: Developing the Mini-Max HB Sampling Method}
This section describes
what practising the checklist looked like on a real design problem, including the
parts that did not go as planned.

The method itself can be put in a sentence. Bethel allocation \cite{bethel1989} searches for
the cheapest multivariate sample that still meets a set of precision
targets; Hierarchical Bayes (HB) small area estimation borrows statistical
strength across regions so that small, data-poor domains can still be
estimated precisely; put the two together and the method hunts for the
smallest sampling fraction at which every target variable clears its
precision target, nationally and in every domain, once HB shrinkage is
taken into account. The full method and results are in \cite{tam2026a}.

What follows is organised around the seven-item checklist in Section~6 above.

\textbf{Item 1: State the target in advance.} Before any AI-assisted
implementation began, the design objective was fixed: the smallest sampling
fraction meeting stated national and domain coefficient-of-variation
targets, for named target variables, at a named geography. Specifically, the
national coefficient-of-variation target was set at 3 per cent and the
domain target at 8 per cent; the design also had to meet stated absolute and
maximum absolute relative error targets (ARE of less than 1\% and MARE of less than 5\% for employment and hours of work, and 45\% for unemployment), a requirement that
actual credible-interval coverage not differ significantly from the nominal
95 per cent, and a convergence diagnostic (multi-chain Gibbs $\hat{R}$)
below 1.05. This is FPOS Principle 2 (professional and scientific method
choice, Table~\ref{tab:fpos}) in direct operation: the target was a
professional judgement made up front, not a description assembled after
seeing what the assistant produced.

\textbf{Item 2: Test against known truth, synthetic and real.} The method
was first tested against a synthetic labour-force population with truth
known by construction. Meeting the stated targets on this population was
not automatic: the hierarchical Bayes prior on the between-stratum random
effect had to be calibrated, and we learnt that a grid search over
candidate prior hyper-parameters was needed before the design met every
stated target. Once calibrated, the design achieved an 80 per cent
sample-size reduction, confirmed by a 1,000-replication Monte Carlo study
reported in \cite{tam2026a} with credible-interval coverage close to the
nominal 95 per cent, and reproduced across four further synthetic
populations. The identical question was then put to real data: a genuine
2021 Australian Census extract, the ABS's confidentialised 5 per cent basic microdata product (the Census's own labour-force-status
question; 55 strata, 8
domains matching Australia's states and territories), treated as a finite pseudo-population: because the extract is a confidentialised sample rather than a full census, ``truth'' here means the known value within that extract, not the true value for the whole Australian population. Checklist item 2 treats
these as the same kind of test, differing only in where the truth comes
from; both are reported, not just the more convenient one. Getting there
took an extra step the synthetic population had not required: we found
that the domain coefficient-of-variation target itself had to be checked
against, and aligned with, an external standard published by the ABS,
described in Item 3. Because a single draw cannot tell you whether a
credible interval is properly calibrated, only whether one point estimate
got lucky, a repeated-sampling coverage check followed: 50 independent
fresh draws from the real population, refitting the model each time,
checking whether the resulting 95 per cent interval actually contained the
truth close to 95 per cent of the time. Nationally, it did: 100 per cent
for employment, 96 per cent for unemployment. The domain-level picture was
not so tidy, as described below.

\textbf{Item 3: Iterate, then disclose.} On the synthetic population, when
the true differences between domains were unusually small, we found
domain-level coverage weakening. We tried a weaker prior, then a
Prasad--Rao correction \cite{prasadrao1990}; neither reliably closed the
gap, and we traced the shortfall to shrinkage bias inherent to how HB
borrows strength across areas, and disclosed it as a residual limitation
rather than dropping the difficult cases from the test suite.

On the real data, our first attempt at the check found no achievable
reduction at all. Rather than accept that, we went back to the precision
target itself: the domain target we had been using, 8 per cent, had never
actually been checked against anything external. We located the ABS's own
published Labour Force methodology, which sets a 25 per cent Relative
Standard Error cut-off for state and territory labour force estimates, reset our target to match it, and reported the change, a concrete
instance of FPOS Principle 9 (use of recognised standards) as much as it is
item 3. Under the corrected target, the same real data and the same method
found a genuine 90 per cent reduction (4,203 respondents from 42,018), with
national point estimates accurate to well under 1 per cent for both target
variables.

The domain-level coverage check on the real data needed more work.
Employment's credible interval covered the truth in only 70 per cent of
replications in New South Wales, and in under 90 per cent in Victoria,
Queensland, and the Northern Territory, well outside what 50 replications
could produce by chance if true coverage really were 95 per cent, even
though unemployment's domain coverage held up, at 90 to 100 per cent
everywhere. New South Wales is the largest domain in the population, so a
small-domain explanation did not fit, and we followed the same iterative
process used throughout this case study: propose an explanation, test it
against the data, discard what does not hold up, and try again.

We first compared the model's own stated uncertainty against the
estimator's actual spread across repeated draws, which ruled out the
credible interval's width as the problem: the ratio of empirical to
modelled standard deviation was 0.94 for New South Wales and 0.93 for
Western Australia, both close to 1. We then checked the classical
hierarchical-model shrinkage factor, that is, how much a domain's estimate
leans on its own sample versus the shared regression prediction, which
ruled out differential shrinkage too, coming back almost identical for both
domains, at about 0.20 to 0.25. What we found differing was the accuracy of
the shared prediction itself: it matched Western Australia's true
employment rate to within half a percentage point but missed New South
Wales's by nearly two, and that mismatch, carried through an otherwise
ordinary degree of shrinkage, reproduced the measured bias in both domains
almost exactly.

We fixed this by giving the regression its own free intercept for each
state, so it was no longer required to fit one national pattern onto every
domain, and re-ran the check against the identical fifty draws that had
found the problem. Domain-level employment coverage moved from a range of
70 to 100 per cent across the eight domains to a tight band of 92 to 98 per
cent; New South Wales rose from 70 to 98 per cent, and its bias shrank by
an order of magnitude and changed sign, consistent with noise rather than a
residual fault. One domain, South Australia, moved the other way, from 100
to 92 per cent, within a fifty-replication coverage estimate's own sampling
noise of roughly three percentage points; we report it rather than drop it,
for the same reason we report the rest of what did not go as planned.

Those fifty draws, however, were the same draws used to diagnose the problem in the first place: the shrinkage comparison, the interval-width check, and the shared-prediction diagnosis that identified the fix had all been carried out by looking at exactly these replications. Confirming the fix against the same draws that motivated it shows that the fix resolves what was observed in those draws; on its own, it does not show that the fix generalises to data the diagnosis never saw, precisely the correlated-error risk this paper's own Section~2.2 warns against, and precisely what checklist Item~2 (test against a data-generating process) requires taking seriously rather than passing over. We therefore re-ran the fixed model a second time, independently: a fresh set of fifty replications drawn with a new random seed never used in diagnosis, fitting the identical domain-dummy model in a second, independently written implementation (Python/PyMC, rather than the R/\texttt{mcmcsae} implementation used throughout the rest of this case study), which is a stronger form of independence than a fresh draw alone would give and addresses checklist Item~4 as much as Item~2. On this fresh, independently implemented set, New South Wales's employment coverage was 43/50 (86 per cent, exact 95 per cent binomial interval 73--94 per cent); the Northern Territory, not previously flagged, was also 43/50 (86 per cent); the remaining six domains ranged from 92 to 100 per cent; nationally, coverage was 48/50 (96 per cent). New South Wales's 86 per cent point estimate is lower than the 98 per cent obtained on the diagnostic draws, consistent with what the correlated-error argument predicts might happen, but its exact 95 per cent binomial interval (73--94 per cent) contains the nominal 95 per cent target, so the result on this independent check is not inconsistent with nominal coverage; with only 50 replications the check cannot rule out a true rate somewhat below 95 per cent either, and the fix should be read as corroborated on genuinely independent grounds rather than as proven exactly, with wider uncertainty than the same-draws figure implied. We report both numbers rather than only the more favourable one.

Writing and running each diagnostic script, reading its result, and setting up the next test took
minutes rather than the days the same sequence of checks would otherwise
have needed; the fix we eventually settled on, a domain-specific intercept
in the regression, is standard modelling practice.

\textbf{Item 4: Apply the same scrutiny to the tooling itself.} Two
different kinds of programming error had to be guarded against, and they
are not equally easy to catch. Syntax errors are the easy ones: the program
simply fails to run, and the failure is immediate and unambiguous. Logical
errors, code that runs without complaint but computes the wrong thing, are
harder, and watching the program execute successfully does not rule them
out.

Our approach was to set out the statistical methodology in full before any
code was written, so that the AI assistant was implementing a specified
method rather than inventing one; in our experience this meant the
programming was generally done correctly, though it did not remove the need
to check. Verifying a routine of this kind against a fully independent,
hand-built dataset for every computation it performs is not feasible; the
volume of arithmetic involved rules that out. What was feasible, and what
we relied on, was checking whether the output made sense against what the
methodology should produce. Checking every line of code directly is only a
realistic option for a statistician who is also a competent programmer;
where that was not practical, we instead used a second, independently
configured AI system to review the implementation against the written
methodology, an application of the Analyst-Inspector pattern named in
Section~2, rather than relying on the system that built the code to also
judge whether it was built correctly.

The synthetic study's own simulation check was also re-run independently,
end to end, in a fresh R session separate from where the method was built,
layered across a single draw, an averaged draw, and individual replicates,
to rule out mistaking sampling noise for a genuine discrepancy. The
real-data check caught something more consequential the same way: an early
version of the verification script itself used a fabricated design effect
and treated a partly synthetic continuous variable as if it were directly
observed real data. Neither was a finding about the method; both were
artefacts of the checking tool, caught and removed before any conclusion was
drawn about the method itself. We asked the same question of the checker
that we asked of the thing being checked.

To be concrete about what ``AI-assisted'' meant in practice: a general-purpose large language model assistant, directed by the author, wrote and revised the R implementation (the HB model calls, the Bethel-allocation set-up, and the Monte Carlo verification scripts) against the methodology specified in advance by the author; a second, separately configured session of the same class of assistant was used for the independent code review described above. That is tool independence in the weak sense (a fresh session, prompted separately, without sight of the first session's reasoning) rather than the stronger senses a fully independent audit would have, such as a different provider, a human reviewer with no role in development, or a differently implemented benchmark. The fresh-draw validation reported below under Item~3, which re-implements the fitting routine in a different language and library rather than only drawing new data, is this paper's attempt at that stronger, implementation-level independence; the case study as a whole should be read as one methodologist's use of AI assistance under the checklist, not as an institutionally independent audit, and the auditor role in Section~2.1 is correspondingly aspirational for a single-author case study of this kind.

All computations were carried out in R using three established statistical
packages: \texttt{mcmcsae} (Boonstra~2021) for HB~model fitting,
\texttt{R2BEAT} (Barcaroli~et~al.~2023) for Bethel allocation, and
\texttt{sae} (Molina \& Marhuenda~2015) for the Prasad--Rao MSE correction.
Data manipulation uses \texttt{data.table} and \texttt{dplyr}.

The implementation follows an \textit{assemble-to-order} rather than
\textit{made-to-order} strategy. In a made-to-order approach, the analyst
implements every algorithm from scratch---writing their own MCMC sampler,
Bethel optimiser, and MSE correction routine---and bears the full burden of
validating each component. In an assemble-to-order approach, pre-validated
components are sourced from specialist developers and orchestrated into a
purpose-built pipeline. The testing obligation is shared rather than eliminated: component correctness is primarily the responsibility of the component developers, while the adopting analyst and institution remain accountable for fitness for purpose, correct parameterisation, version control, and validation of the complete production pipeline built around those components.

Each component used here has published, peer-reviewed or otherwise externally documented validation evidence, a level of independent scrutiny bespoke code rarely receives, though it is evidence about the component in general rather than a guarantee for this specific use. \texttt{mcmcsae} is developed and maintained
by Statistics~Netherlands and includes a built-in Simulation-Based
Calibration routine (\texttt{SBC\_test}) that formally verifies sampler
correctness against known posteriors---a standard that goes beyond ordinary
unit testing. \texttt{R2BEAT} is developed by Istat and peer-reviewed in
\textit{The R Journal} (2023). \texttt{sae} is peer-reviewed in \textit{The
R Journal} (2015) and is the standard reference implementation of the
Fay--Herriot model \cite{fayherriot1979} and Prasad--Rao correction. The MC simulation pipeline
assembles these validated components; it does not reimplement them. The
coverage results reported here therefore rest on a tested computational
foundation that would not have been available under a made-to-order
approach.

\textbf{Item 5: Test at the real operating scale.} Once the domain target
was corrected to the real ABS standard, we ran the check at a realistic
sample size, a 5 per cent draw matching the scale an actual collection
would use, rather than an inflated one chosen to make a reduction easier to
find. The 90 per cent figure above is this real-scale result.

\textbf{Item 6: Prefer synthetic or aggregate data; treat real microdata
as a governed exception.} This is the second pillar in direct operation. We
used synthetic populations as the default throughout, built so that no
confidential respondent record needed to pass through any tool, AI-assisted
or otherwise, for most of the verification work. The one real-data check in
this case study used a public-use Census microdata extract released under
its own governed access arrangement, handled under the same controls that
would apply regardless of which tool performed the computation: a
case-by-case exception, not a default.

\textbf{Item 7: Keep a dated record.} We kept each stage of this case
study, the original synthetic result, its independent re-verification, the
real-data check, the correction to the ABS-sourced target, and the re-run
that followed, as a separate, dated script with its own header explaining
what changed and why, rather than a single undocumented final number. That
record is what makes every claim in this section checkable by someone who
was not in the room when it was made.

This section reports the case study at the level the checklist requires: targets, tests, and results. The full mathematical specification of the Bethel/HB Mini-Max method, the complete simulation design, and the full set of domain-level result tables are given in the companion methodological paper \cite{tam2026a}, to which the reader is referred for those details.

This is what the checklist looked like when applied to one real design
problem: a target set out in advance in Item 1, checked in Items 2 and 4
against evidence we did not build, iterated on in Item 3 when it
fell short, and kept clear of respondent data everywhere Item 6 did not
require otherwise. AI assistance changed how quickly we could repeat Items
2 and 4; it did not change what those items required us to do.

\section{Conclusion}
Responsible use of AI in official statistics is not primarily a question of
how much of a method or its implementation was produced by a human versus an
algorithm. It rests on two things: whether an independent verification
protocol stood between the method and its adoption for production use, and
whether respondent confidentiality was protected throughout.

This paper has set out both, and shown the verification protocol at work in developing the
Mini-Max HB sampling methodology. On synthetic data built to
resemble a real survey: an 80 per cent reduction in required sample
size, verified by Monte Carlo simulation against independently checkable
benchmarks. On 2021 Census microdata: a precision target corrected
to match the ABS's own published reliability standard enabled a 90 per cent
reduction with national accuracy well under 1 per cent. Because the
protocol asks for disclosure rather than a tidy ending, a domain-level
coverage failure in the largest domain in the population that did not fit
the explanation offered for smaller ones was followed through rather than
left as a loose end: ruled out as an interval-width problem, ruled out as
differential shrinkage, traced to a shared regression model that fit every
domain but one, and corrected with a domain-specific intercept. Because confirming that fix only against the same replications that had exposed the problem would itself be a correlated-error risk, we validated it a second time against a fresh set of replications, refit in an independently written implementation: coverage for that domain rose from 70 per cent under the original model to 86 per cent on this independent check (exact 95 per cent binomial interval 73--94 per cent) --- a result that is not inconsistent with nominal 95 per cent coverage, and a genuine improvement corroborated on independent grounds, though 50 replications alone cannot rule out a true rate somewhat below 95 per cent, reported alongside, rather than instead of, the 98 per cent obtained on the diagnostic draws themselves.

What matters more than the numbers themselves is what produced them: a
target stated before testing began, a check run against something the
assistant did not build, an honest return to the evidence when the first
answer looked too clean, a diagnosis reached through repeated rounds of
testing and retesting, sped up by AI-assisted programming and resolved with
standard statistical practice, and respondent
data kept out of the loop everywhere the check did not need it. The
specific illustration is one experience. The protocol --- methodologist, assistant, auditor, and disciplined data protection throughout --- is what is meant to transfer to any national statistical office weighing how to adopt AI responsibly, whether or not its own results turn out as tidily.

Two things this paper does not do are worth stating plainly rather than leaving implicit. First, as Section~1 noted, this protocol is a narrow, operational instrument for the validity-and-robustness and privacy-and-security principles specifically in AI-assisted methodology development; it is not a substitute for a comprehensive responsible-AI framework such as UNECE's~\cite{unece2025}, which also covers ethical purpose, accountability, transparency, and fairness across the full statistical production cycle, well beyond the moment of methodology verification this paper addresses. Second, passing this protocol's checklist verifies and validates a method up to the point of production adoption; it says nothing about the method's behaviour afterwards. A method that met every target in testing can still drift once it meets real, changing production data, which is why ongoing monitoring after deployment, checking that production performance continues to track what verification predicted, is a necessary sequel to this protocol, not an optional extra, even though building out that monitoring regime is outside this paper's scope.

\section*{Acknowledgments}
The author thanks the two anonymous referees for their helpful comments on an earlier version of this paper.
The algorithms described in this paper --- the R implementation of the
Mini-Max HB sampling method, its Monte Carlo verification scripts, and the
independent Python/PyMC re-implementation used for the fresh-draw
validation in Section~7 --- were prepared with Claude (Anthropic), directed
by the author, and were themselves subject to the verification and
confidentiality protocols set out in this paper. Claude was also used to
improve the flow and expression of the writing throughout the paper. The
author takes full responsibility for the content.

\end{document}